\documentclass[a4paper,11pt]{article}
\pdfoutput=1
\usepackage{jcappub}

\usepackage{graphicx}
\usepackage{dcolumn}
\usepackage{bm}
\usepackage{multirow}
\usepackage[dvipsnames]{xcolor}
\usepackage[pscoord]{eso-pic}
\usepackage[normalem]{ulem}
\usepackage{slashed}
\usepackage{makecell}
\usepackage{upgreek}
\usepackage{listings}
\usepackage{comment}

\newcommand{\placetextbox}[3]{
	\setbox0=\hbox{#3}
	\AddToShipoutPictureFG*{
		\put(\LenToUnit{#1\paperwidth},\LenToUnit{#2\paperheight}){\vtop{{\null}\makebox[0pt][c]{#3}}}
	}
}

\newcommand{\eqsp}{\,}

\newcommand{\nub}{{\lower-0.8ex\hbox{\textbf{\fontsize{3pt}{3pt}\selectfont(---)}}\mkern-10.5mu \nu}}

\newcommand{\tquote}[1]{``#1''}

\author[1]{Thomas Hambye,}
\author[1]{Marco Hufnagel,}
\author[1]{and Matteo Lucca}
\affiliation[1]{Service de Physique Th\'{e}orique, Universit\'{e} Libre de Bruxelles, C.P. 225, B-1050 Brussels, Belgium}

\emailAdd{thomas.hambye@ulb.be, marco.hufnagel@ulb.be, matteo.lucca@ulb.be} 

\title{Cosmological constraints on the decay of heavy relics into neutrinos}

\abstract{
	A massive particle decaying into neutrinos in the early Universe is known to be less constrained than if it was decaying into other standard model particles. However, even if the decay proceeds into neutrinos, the latter still inevitably emit secondary particles undergoing electromagnetic interactions that can be probed. We analyse in details how sensitive various cosmological probes are to such secondary particles, namely CMB anisotropies, CMB spectral distortions, and Big Bang Nucleosynthesis. For relics whose lifetime is shorter than the age of the Universe, this leads to original and stringent bounds on the particle's lifetime as a function of its abundance and mass.
}

\begin{document}

\vspace*{0.1 cm}
\placetextbox{0.90}{0.97}{\small ULB-TH/21-21}

\maketitle

\section{Introduction}\label{sec: intro}
Many beyond the standard model (BSM) scenarios imply the presence of new unstable particles during the early Universe epoch. For instance, dark matter (DM) setups introducing new dark sector interactions typically predict the existence of additional mediator particles, whose fate can affect the whole thermal history of the Universe, spanning from Big Bang Nucleosynthesis (BBN) predictions and the shape of the Cosmic Microwave Background (CMB) anisotropy power spectra to the cosmic rays and high-energy neutrino spectra observed today. Indeed, this is precisely the case for a wide range of models, such as for instance self-interacting DM scenarios involving the exchange of a light mediator~\cite{Feng:2009hw,Buckley:2009in,Loeb:2010gj,Tulin:2013teo,Bringmann:2016din, Hambye:2019tjt} or hidden sector DM setups communicating with the SM through a portal interaction.

If these unstable particles have still not disappeared by today, they can be probed
by direct and indirect detection and also have a cosmological impact, such as having a too large relic abundance, thereby overclosing the Universe (unless they have tiny sub-eV masses or their numbers are severely suppressed at some point). If instead they do decay with a lifetime smaller than the age of the Universe, their impact, which depends on their characteristics (such as mass and primordial abundance), is mostly cosmological.

Most cosmological probes are basically only sensitive to the amount of electromagnetic (EM) material that is effectively injected into the early Universe photon-baryon fluid (see e.g.~\cite{Slatyer:2012yq, Poulin2017Cosmological}). Thus, one could at first sight expect that if an extra particle, henceforth generically referred to as ``$\phi$'', decays dominantly into non-EM material, such as into neutrinos or dark radiation, the decaying particle will be poorly constrained by these probes. However, even if the decay proceeds into neutrinos, e.g.~via $\phi \rightarrow \nu \bar{\nu}$ or $\phi \rightarrow \nu {\nu}$,  still, charged and colored particles will be produced inevitably via the emission of additional on-shell or off-shell $W/Z$-bosons. This ultimately results in the production of EM and hadronic showers of particles that can affect electromagnetically the CMB and BBN. This way, by carefully evaluating the amount of EM material that the neutrinos release, the decay of massive particles into neutrinos can be constrained by cosmological probes. To determine these limits constitutes the main goal of this work.

In the following, we will focus our analysis on the cosmological constraints coming from CMB anisotropies, CMB spectral distortions (SDs), and BBN photodisintegration. We will do that for the case where the particles have a lifetime shorter than the age of the Universe. This implicitly implies that the massive particle we assume to be decaying cannot constitute the DM observed today, but still it can play a role in the early Universe. Complementary constraints that rely on the assumption that the decaying particle is instead the DM, and hence constrain lifetimes larger than the age of the Universe, can be found in e.g.~\cite{Esmaili:2012us, ElAisati:2015ugc, Slatyer:2016qyl, Garcia-Cely:2017oco, IceCube:2018tkk, Bhattacharya:2019ucd, Coy:2020wxp, Coy:2021sse}.

In light of these considerations, our analysis is organized as follows. In Sec.~\ref{sec:injection} we will outline the method we employed to derive the fractional amount of energy emitted in the form of EM radiation by the decay process and present the resulting relation between this quantity and the relic's mass. In Sec.~\ref{sec:probes} we will briefly describe the cosmological probes we will use to constrain the characteristics of the decay. In Sec.~\ref{sec:dis} we will combine the two previous sections to derive original bounds on the decay of massive relics into neutrinos. We close in Sec.~\ref{sec: sum} with a summary of our results and other concluding remarks.

\section{Energy injection}\label{sec:injection}
To calculate the amount of EM energy (i.e.~carried by photons and charged particles) that is produced by the decay of a massive relic into primary neutrinos, in the following we have to \textit{(i)} model the particle shower production, and \textit{(ii)} quantify the interactions of the shower products with the background plasma. We thereby explicitly focus on decays into electron neutrinos, but note that any other flavour would give similar results (due to lepton universality of electroweak interactions).

\subsection{Electroweak shower}
The emission of secondary particles from neutrinos through radiative corrections has already been considered in various contexts in the literature (see e.g.~\cite{Cirelli2011PPPC,Queiroz:2016zwd,Slatyer:2016qyl,Stocker:2018avm,Liu:2019bbm}). These studies usually rely on \textsc{Pythia}8~\cite{Sjostrand:2014zea} to properly simulate the electroweak shower. This approach assumes a factorization between the process producing the primary neutrinos and the gauge-boson emission. This is justified for neutrino energies above the mass of the $W/Z$ bosons, in which case the dominant contribution comes from the emission of one (or multiple) on-shell $W/Z$-bosons from any of the final-state neutrinos. For a decay, this applies when $m_\phi\gtrsim 2\,m_{W,Z}$. For these masses, the amount of EM energy that gets injected into the plasma becomes effectively independent of the underlying BSM model and we can follow the established procedure to simulate the shower by \textit{(i)}~generating spherically symmetric $\nu_e \bar{\nu}_e$ events with energy $E_{\nu_e} = E_{\bar{\nu}_e} = m_\phi/2$, and \textit{(ii)}~using these events as an input for \textsc{Pythia8.3}~\cite{Sjostrand:2014zea} with the electroweak shower option enabled.\footnote{Specifically, we use the two options \texttt{PartonShowers:model=2} and \texttt{Vincia:EWmode=3}.}

For smaller masses, $m_\phi/2 \lesssim m_{W,Z}$, the primary decay and the emission process do no longer factorize, meaning that \textsc{Pythia}8.3 alone is no longer able to simulate the production of secondary particles. In fact, using the same procedure as before for these masses, \textsc{Pythia}8.3 predicts that no EM material is injected at all (see~Fig.~\ref{fig:zeta} below).\footnote{Similarly, the PPPC results~\cite{Cirelli2011PPPC}, which have been obtained with \textsc{Pythia}8.1, are to be used only for neutrino energies above $100-150\,\mathrm{GeV}$.} Nevertheless, for $m_{W,Z} < m_\phi < 2\,m_{W,Z}$ an on-shell gauge boson emission is still possible from a neutrino with energy $E_\nu > m_\phi/2$, and for $m_\phi<m_{W,Z}$ the gauge boson can still be produced off-shell. In these cases, the dominant process is $\phi\rightarrow \nu_e f_1 f_2 f_3$ with any allowed combination of three SM fermions $f_{1,2,3}$. To take this process into account, we instead have to consider the full transition amplitude for the emission process, which ultimately makes the calculation dependent on the actual BSM scenario.

In order to quantify this effect, we consider a simple scenario where the decaying particle is a massive vector boson coupling to neutrinos in the following simple way
\begin{align}
    \mathcal{L} \supset \lambda_e \phi_\mu \bar{\nu}_e \gamma^\mu P_L \nu_e\eqsp.
    \label{vectorcoupl}
\end{align}
Strictly speaking, in its simplest version, this model also predicts the production of a pair of charged leptons via $SU(2)_L$ invariance, which will dominate the production of EM material. Nevertheless, without entering here in the details of specific models, the amount of EM energy obtained from neutrinos will still be generic of what one could get in other models not necessarily producing charged leptons. Note, however, that Eq.~(\ref{vectorcoupl}) can be obtained without a $SU(2)_L$ charged lepton counterpart from the decay of a vector boson into two heavy sterile neutrinos, mixing with ordinary neutrinos, in the spirit of Ref.~\cite{Coy:2020wxp}. One could also consider frameworks based on the existence of a scalar triplet. For instance, the neutral component of a scalar triplet, which feebly couples to doublets of leptons, decays only into two neutrinos, whereas its singly and doubly charged $SU(2)_L$ partners could have disappeared long before, e.g.~by decaying more quickly via gauge interactions (such as through the process $\Delta^+\rightarrow \Delta^0 l \nu$). One can check that this example of a scalar particle decay leads to a production of particles that is very similar to the one obtained from the decay of a vector boson, Eq.~(\ref{vectorcoupl}). In this way, the results obtained in this case should be rather general and reflect what one can expect also in other scenarios.

By implementing this model in \textsc{FeynRules}~v2.3~\cite{Alloul:2013bka} and afterwards using it inside of \textsc{MadGraph5\_aMC@NLO}~v3.2.0~\cite{Alwall:2011uj}, we then fully simulate events for the processes $\phi \rightarrow \nu_e f_1 f_2 f_3$ and $\phi \rightarrow \bar{\nu}_e f_1 f_2 f_3$, where $f_{1,2,3}$ can be any combination of physically allowed SM fermions. The so-generated events are then used as in input for \textsc{Pythia}8.3, which takes care of the remaining parts of the shower process, including photon radiation and hadronization.

For the full range of masses, we then end up with a set of shower products including stable particles ($e^\pm$, $\gamma$, $\nu_i/\bar{\nu}_i$, $p^\pm$) as well as particles which can decay through weak interactions ($\mu^\pm$, $\pi^\pm$, $K^\pm$, $n/\bar{n}$). Once produced, all of these particles can still undergo interactions with the background plasma, which we discuss in the following.

\subsection{Interactions with the background plasma}\label{sec: inter bg}

The characteristic timescales for \textit{(i)} an unstable particle to decay via weak interactions and \textit{(ii)} a particle of the shower to interact electromagnetically with the background plasma are very similar. Consequently, these two effects are intertwined, meaning that in order to obtain the actual fraction of energy that is injected into the EM plasma, it is usually necessary to solve the full set of Boltzmann equations for the different shower products. In the following we show, however, that the scattering reactions do not lead to a significant redistribution of energy, which is why they can essentially be neglected.

Photons and electrons from the shower interact with the background plasma mainly via double-photon pair creation $\gamma \gamma_\text{th} \rightarrow f \bar{f}$ and inverse Compton scattering $e^\pm \gamma_\text{th} \rightarrow e^\pm \gamma$ ~\cite{Kawasaki:1994sc}. Here, $f$ is a SM fermion and the index 'th' indicates particles from the thermal background. In particular, the pair production of electrons and positrons will occur if $E_0 \gtrsim m_e^2/[22T(\tau_\phi)]$~\cite{Kawasaki:1994sc}, i.e.~if $m_\phi/2\gtrsim E_0$, which we find to be true for all parameter combinations considered in this work. In this case, the given processes conserve the energy in the EM sector. However, for $m_\phi/2 > m_\mu^2/[22T(\tau_\phi)]$, the process $\gamma \gamma_\text{th} \rightarrow \mu^- \mu^+$ becomes relevant. Unlike the electron pair production process, this reaction will transfer back part of the energy to the neutrino sector via the muon decay $\mu^- \rightarrow e^- \bar{\nu}_e \nu_\mu$.
By comparing the reaction rates for the pair production of electrons with the one for muons for an average photon energy $\bar{E}_\gamma \sim 10^{-2}m_\phi$ and at a temperature $T(\tau_\phi)$, we find that the energy loss via muon production leads to a smaller than $1\%$ correction for $\tau_\phi \gtrsim 10^{4}\,\mathrm{s} \times (m_\phi/\mathrm{PeV})^{2}$. This covers almost the full parameter space considered in this work (except for the shaded region in Fig.~\ref{fig:final} below). Thus we neglect this effect for simplicity.

A similar reaction can also be induced by neutrinos, i.e.~$\nu_i \bar{\nu}_{i, \text{th}} \rightarrow f \bar{f}$, which can instead transfer energy from the neutrino sector to the EM sector~\cite{Acharya:2020gfh}. However, this process is only relevant if the intermediate $Z$ boson is produced close to resonance. This "Glashow-like resonance" happens roughly when $m_\phi/2 \sim E_\text{GR} \sim m_Z^2/[\mathcal{O}(1)T_\nu]$, which we deduce by solving the full Boltzmann equation for this process. Hence, this process is relevant only if $m_\phi > 10^7\,\mathrm{GeV}$ for $\tau_\phi > 10^3\,\mathrm{s}$. Coincidentally, \textsc{Pythia}8.3 only works reliably for $E_0 < 10^7\,\mathrm{GeV}$, which is why we only consider energies below this (artificial) threshold, meaning that pair production from neutrinos can also be neglected. A similar argument can also be made for reactions of the form $\nu_i e^-_\text{th}\rightarrow \dots$, which feature a slightly more accessible resonance at $E_\text{GR} \sim m_W^2/m_e$, but are also additionally suppressed by the small number density of thermal electrons.

Finally, the particles $x\!\in\!\{\mu^\pm, \pi^\pm, K^\pm, p^\pm, n/\bar{n}\}$ can induce additional energy injection via \textit{(i)} their decay and/or \textit{(ii)} scattering reactions~\cite{Kawasaki:2004qu,Jedamzik:2006xz} such as Thompson scattering $x \gamma_\text{th} \rightarrow x \gamma$. However, as already mentioned above, both of these effects are potentially intertwined due to their similar timescales. For example, pions mainly decay via $\pi^- \rightarrow \mu^- \bar{\nu}_\mu \rightarrow e^- \bar{\nu}_e \nu_\mu \bar{\nu}_\mu$, and thus deploy most of their energy into the neutrino sector. However, the scattering reactions prior to the decay instead transfer energy from the pions into the EM plasma, which can potentially distort the $1:3$ ratio implied by the decay alone. To quantify this effect, we approximate the energy loss as continuous via a differential equation of the form $\dot{E}_x = D_x(E, t)$ with the energy-loss rate $D_x$ of $x$,\footnote{For a list of the relevant reactions and their corresponding energy-loss rates, see e.g.\ App.~B of~\cite{Kawasaki:2004qu}.} which we solve from an initial time $t_i=\tau_\phi$ until some final time $t_f$ satisfying $t_f = \tau_\phi + \min\{E_x(t_f)\tau_x/m_x, t_\text{pdi}\}$, where $t_\text{pdi}$ is the typical timescale at which photodisintegration reactions become important. The transferred energy $\Delta E_x = E_x(t_i) - E_x(t_f)$ is then attributed to the total amount of EM energy, while the remaining decay of the particle $x$ with energy $E_x(t_f)$ is properly handled by running \textsc{Pythia}8.3 again. Overall we find that the inclusion of this energy-loss mechanism does not significantly change our results, as it merely leads to  $\sim1\%$ corrections for $\tau_\phi > 10^4\,\mathrm{s}$. Consequently, it is justified to neglect this effect altogether and simply let all unstable particles decay immediately, which can be handled in the same \textsc{Pythia}8.3 run as the shower generation itself.

\subsection{Results}
\begin{figure}[t]
    \centering
    \includegraphics[width=0.55\columnwidth]{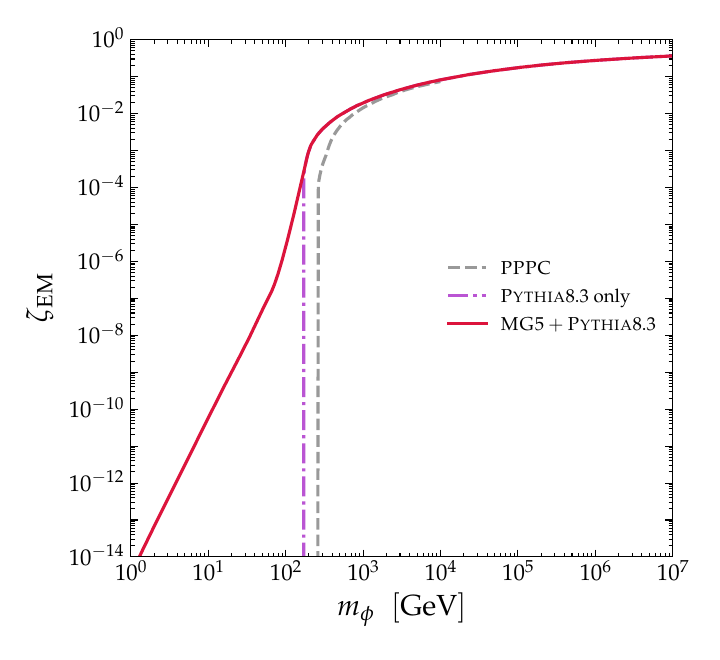}
    \caption{The fraction $\zeta_\text{EM}$ of EM energy that is injected into the plasma as a function of the mass $m_\phi$ (solid red). For comparison, we also show the results that we obtain when using only \textsc{Pythia}8.3 (dash-dotted purple), as well as the results obtained with the PPPC (dashed gray).}
    \label{fig:zeta}
\end{figure}

Having described the different steps of our calculation, in Fig.~\ref{fig:zeta} we show the fraction $\zeta_\text{EM}$ of energy injected into the EM plasma as a function of the mass $m_\phi$ (solid red). In addition, we also show the results that we obtain when only using spherically symmetric $\nu_e \bar{\nu}_e$ events as an input for \textsc{Pythia}8.3 (dash-dotted purple), as well as a comparison with the previous results from the PPPC~\cite{Cirelli2011PPPC} (dashed gray).

We find that our results agree well with both other approaches for large masses. However, for $m_\phi$ around and below $\sim 2 \,m_{W,Z}$, the three curves start to differ substantially. The deviation between the \textsc{Pythia}8.3 and PPPC curves for $m_\phi \sim 2\,m_{W,Z}$ can be explained by noting that the PPPC is based on \textsc{Pythia}8.1, which was subject to a bug preventing it from producing correct results in this region of parameter space (see~\cite{Bauer:2020jay} for further details) and which has since been fixed in \textsc{Pythia}8.3.\footnote{We explicitly checked that \textsc{Pythia}8.3 can reproduce the results from~\cite{Bauer:2020jay}, which used an independent technique to properly simulate the shower.} For even smaller masses, below $m_\phi \sim 2 \, m_{W,Z}$, \textsc{Pythia}8.3 itself is no longer able to simulate the initial gauge boson emission from spherically symmetric events, and instead it becomes essential to properly generate the initial events with \textsc{MadGraph5\_aMC@NLO} as explained above. In fact, while $\zeta_{\text{EM}}$ would simply drop to zero at $m_\phi \sim 2\,m_{W,Z}$ when relying solely on \textsc{Pythia}8.3, a proper treatment instead leads to a sizeable production -- albeit slightly model dependent --  even for values of $m_{\phi}$ several orders of magnitude below the $W/Z$ thresholds. More precisely, we find that $\zeta_\text{EM}$ does not simply drop to zero at $m_\phi \lesssim 2 \,m_{W,Z}$, but instead falls off continuously with $\zeta_\text{EM} \propto (m_\phi/m_W)^4$ for $m_\phi < m_{W,Z}$. The latter expression is indeed expected from the off-shell behavior of the gauge-boson propagator in the transition amplitude. 

Overall, the results presented in this work do not only correct the previous literature regarding this matter for $m_\phi \sim 2 \, m_{W,Z}$, but also extend it to even smaller masses. We will see that, despite this Fermi suppression, the resulting particle production can still have cosmological effects that can be probed for masses as low as $\sim$~GeV. It is to be anticipated that this could also have effects in other contexts, such as for DM indirect detection, and modify at low energies various results accordingly.

\section{Cosmological probes}\label{sec:probes}
Now that the properties of the energy injection caused by the decay of the massive relic have been outlined, we move on to the description of the cosmological probes we will use to constrain such a process. In particular, we will discuss CMB anisotropy and SD constraints in Secs.~\ref{sec: anis} and \ref{sec: SDs}, respectively, as well as BBN bounds in Sec.~\ref{sec: BBN}. For each case we will briefly describe the main effects that non-standard decays have on that given probe, review existing constraints together with their limitations, and discuss how to combine them with the results obtained in Sec.~\ref{sec:injection} for the specifics of the decay process considered here.

\subsection{CMB anisotropies}\label{sec: anis}
As it is very well known in the literature (see for instance~\cite{Galli2013Systematic, Slatyer2015Indirect} for detailed reviews), the main impact that an energy injection into the CMB photon fluid has on the CMB anisotropy power spectra is given by the inevitable modification it causes to the recombination and reionization evolution. This happens on the one hand because of the increased photon (and baryon) temperature, which can alter the thermal history of the Universe (and hence the evolution of the free electron fraction), and on the other hand because the injection of additional high-energy particles can directly re-ionize the plasma if recombination has already happened.

In the context of decaying massive relics, the relative contribution of these effects and their overall impact on the CMB power spectra is mainly dictated by the lifetime $\tau_\phi$ of the decaying particle and by the total amount of energy that is effectively injected in the form of EM radiation,\footnote{In full generality, with EM radiation here we mean electrons, positrons, and photons. However, since the energy-loss mechanism for positrons is the same as for electrons, with the additional possibility to annihilate with the thermal electrons to produce two photons, in this setup a positron is nothing else than an electron and two photons. Furthermore, for initial injection energies above $\mathcal{O}$(GeV), which are the ones relevant in this work (see Fig.~\ref{fig:zeta}), the injections of electrons and photons quickly translate into one another because of the very efficient scattering and pair production processes (see e.g.~Sec.~2 of~\cite{Acharya2019CMB}). Therefore, for this probe the injection of either of these three particles, electrons, positrons and photons, is almost perfectly interchangeable (as it becomes clear e.g.~from Fig.~9 of~\cite{Acharya2019CMB} for such high energies) and we will thus only generically refer to EM radiation.} i.e.~by the particle's mass $m_\phi$, its fractional abundance $f_{\rm \phi} = \Omega_\phi/\Omega_\text{DM}$, and the amount $\zeta_\text{EM}$ of that fraction that gets deposited into the plasma in the form of EM radiation (see e.g.~\cite{Poulin2016Fresh, Poulin2017Cosmological} for more details). Here, $\Omega_\phi$ corresponds to the abundance that the $\phi$ particle would have today if it was not decaying. Therefore, given the particle's lifetime, it is possible to calculate the maximum amount of energy that can be injected into the system without spoiling the goodness of the fit to CMB anisotropy data. Typically, this will then result in exclusion bounds in the $f_{\rm \phi}\zeta_\text{EM} - \tau_\phi$ plane, with a mild dependence on the particle's mass displayed in form of a width on the final limits~\cite{Slatyer:2016qyl, Poulin2017Cosmological, Acharya2019CMB}.

Here we adopt the recent bounds presented in~\cite{Acharya2019CMB}, based on Planck 2015 data \cite{Ade2015PlanckXIII}.\footnote{To our knowledge, the only similar bounds derived using Planck 2018 data \cite{Aghanim2018PlanckVI} are those reported in~\cite{Lucca2019Synergy}. However, they are based on a mock likelihood and assume a more simplified injection history with respect to~\cite{Acharya2019CMB}. We decided therefore to employ the latter, and in doing so we are in any case conservative.} They coincide with the previous literature~\cite{Slatyer:2016qyl, Poulin2017Cosmological} for lifetimes larger than recombination, but are improved for smaller lifetimes. These improvements result from more carefully following the delay between injection and deposition time of the energy, thus extending the constraints down to redshifts of the order of $z\simeq 10^4$. In particular, since the mass range where the CMB anisotropy bounds are relevant in our case is always well above 1$\,\mathrm{GeV}$ (even assuming $f_\phi$ as large as $10^2$, see Fig.~\ref{fig:final}), here we will make use of the bounds derived for 1 TeV particles in Fig.~9 of the reference. This approximation is possible and quite accurate as this bound is almost identical to the one presented for 1$\,\mathrm{GeV}$ injections in the same figure.

\subsection{CMB spectral distortions}\label{sec: SDs}
Because of the aforementioned effects, CMB anisotropy constraints play a dominant role when a significant fraction of the baryons in the Universe is ionized. However, before recombination, when the free electron fraction is equal to one, CMB anisotropies quickly become insensitive to eventual energy releases and other cosmological probes need to be considered. The first example we will discuss is given by CMB SDs (see e.g.~\cite{Chluba2019Spectral, Chluba2019Voyage, Lucca2019Synergy, Fu2020Unlocking} for recent reviews of the topic).

CMB SDs are, generally speaking, any type of distortion of the CMB photon spectrum from a pure black body shape. They are caused by the gradual inefficiency of number changing and scattering processes (such as Bremsstrahlung and Compton scattering) and the consequent inability to restore the black body spectrum after an injection of additional EM energy. These are predicted to exist even within the $\Lambda$CDM paradigm~\cite{Chluba2011Evolution, Chluba2016Which}, and are of course generated whenever extra energy is injected into the photon plasma via, for instance, the decay of massive relics~\cite{Hu1993thermalizationII, Chluba2014Teasing, Acharya2019New, Lucca2019Synergy, Chluba:2020oip, Acharya:2021zhq}.

For this reason, SDs are \textit{per se} a very powerful constraining tool, which is however limited by the currently outdated observational status. Indeed, although several SD missions have been proposed in the meanwhile~\cite{Kogut2011Primordial, Andre2014Prism, Kogut2019CMB, Chluba2019Voyage}, the most stringent bounds date back to the '90 and the COBE/FIRAS mission~\cite{Mather1994Measurement, Fixsen1996Cosmic}. Nevertheless, despite this limitation, since -- in the context of decaying massive relics -- SDs are sensitive to lifetimes in the range between $\tau_\phi\simeq 10^6\,\mathrm{s}$ and recombination\footnote{In principle, they are still constraining even today but for lifetimes larger than $\tau_\phi\simeq 10^{13}\,\mathrm{s}$ CMB anisotropy constraints inevitably start to dominate.} (see Fig.~\ref{fig:final}), they can still be very important to bridge the gap between BBN and CMB anisotropy constraints. Furthermore, looking towards the future, it is to be expected that SDs will be able to surpass BBN bounds by several orders of magnitude and dominate the constraints in the aforementioned time range, as forecasted in e.g.~\cite{Lucca2019Synergy, Chluba:2020oip, Fu2020Unlocking}.

Concretely, to account for SDs we will consider a combination of the bounds derived in~\cite{Poulin2017Cosmological} and~\cite{Chluba:2020oip} (see Fig.~5 and 19 therein, respectively). The latter reference, which focused on accurately describing the thermalization process in the case of high energy injections, provides the currently\footnote{During the final stages of this work, an updated analysis of these constraints has been put forward \cite{Acharya:2021zhq}. We expect that the difference between the two limits does not affect our results and conclusions significantly.} most accurate SD bounds on the decay of massive particles for small lifetimes ($\tau_\phi<10^7\,\mathrm{s}$). However, since in~\cite{Chluba:2020oip} the authors mainly focused on early times (i.e.~only $\mu$ distortions, as the contribution from $y$ distortions is unaffected by their treatment with respect to the previous literature), here we complement them with the $y$ distortions bounds from~\cite{Poulin2017Cosmological}, which cover instead larger $\tau_\phi$ values. 

Note finally that, unlike in the case of the aforementioned CMB anisotropy limits, the bounds obtained by~\cite{Chluba:2020oip} are expressed in terms of the yield so that their mass dependence can be taken into account without the need for any approximation. On the other hand, those by~\cite{Poulin2017Cosmological} are expressed in the same plane as the CMB anisotropy constraints, which means that one cannot extract exactly the mass dependence of the $y$ bounds in this case. However, as can be seen from Fig.~5 of~\cite{Acharya2019New}, for high particle masses (above 10$\,\mathrm{GeV}$) this dependence significantly reduces, so that the constraints become almost universal and we can adopt them as such.

\subsection{BBN photodisintegration}\label{sec: BBN}
In order to push our bounds to even smaller lifetimes, we also account for constraints from BBN (see e.g.~\cite{Cyburt:2015mya} for a review of the topic). In a nutshell, BBN describes the process of light-element formation in the early Universe, which took place at temperatures of the order of $\mathrm{keV}-\mathrm{MeV}$ (or equivalently redshifts around $z\simeq 10^6-10^9$). Theoretically, BBN predictions follow from a system of Boltzmann equations that describe the various abundances of the involved elements, which can then be probed against observations of such abundances in the first galaxies. This comparison shows a remarkable agreement between the values predicted in the SM and the ones inferred from observations~\cite{10.1093/ptep/ptaa104}.

Consequently, any BSM scenario must not distort the SM abundances by too much. This can in fact be challenging, since any BSM particle can change BBN in many different ways~\cite{Depta:2019lbe,Depta:2020zbh,Hufnagel:2017dgo,Hufnagel:2018bjp,Depta:2020wmr,Kawasaki:2020qxm,Poulin:2015opa,Poulin:2015woa,Forestell:2018txr,Jedamzik:2009uy,Boyarsky:2021yoh}, namely by \textit{(i)} increasing the Hubble rate and thus the expansion rate of the Universe (which is similar to a bound on $\Delta N_\text{eff}$), \textit{(ii)} changing the time-temperature relation via the injection of EM energy, and \textit{(iii)} inducing late-time photodisintegration reactions that can destroy the light nuclei previously created during BBN.  In this work, we focus on rather small abundances of the decaying relics (for larger abundances see~\cite{Boyarsky:2021yoh}), in which case the first two effects are subdominant. However, photodisintegration is known to constrain even very small abundances, ten orders of magnitude below the photon abundance~\cite{Hufnagel:2018bjp,Depta:2020wmr,Kawasaki:2020qxm,Poulin:2015opa,Poulin:2015woa,Forestell:2018txr}. Therefore, this effect needs to be properly taken into account for our analysis. 

To this end, we deploy the public code \textsc{ACROPOLIS}~\cite{Depta:2020mhj}, which allows to calculate the evolution of the light-element abundances due to photodisintegration reactions for any BSM model. In our scenario, these disintegration reactions are specifically induced by the EM particles originating from the electroweak shower, and its associated EM cascade creating a set of non-thermal photons.\footnote{For the energy considered in this work, the non-thermal photon spectrum always takes the universal form~\cite{Kawasaki:1994sc}, with a normalisation proportional to the faction of EM energy that is injected by the shower.} These photons afterwards destroy part of the previously created nuclei, thus potentially leading to conflicts with observations.\footnote{Note that disintegration reactions induced by the high-energy neutrinos, e.g.~$\nu N \rightarrow \dots$ with some nucleus $N$, are subdominant since $\sigma_{\nu N}(E_\nu < 10^7\,\mathrm{GeV}) \lesssim 10^{-33}\,\mathrm{cm}^2$~\cite{Gandhi:1998ri}, while $\sigma_{\gamma N}(E_\gamma \sim \mathrm{MeV}-\mathrm{GeV})\sim 10^{-28}\,\mathrm{cm}^2$~\cite{Poulin:2015opa}.~Also reactions of the form $e^\pm N \rightarrow \dots$ can be neglected, since the non-thermal electron/positron spectrum $f_{e^\pm}$ is strongly suppressed compared to the non-thermal photon spectrum $f_\gamma$, i.e.~$f_{e^\pm} \ll f_\gamma$~\cite{Hufnagel:2018bjp}.} We then calculate the corresponding constraints by comparing the light-element abundances after disintegration with the most recent measurements for deuterium ($\text{D}/{}^1\text{H} = (2.547 \pm 0.035)\times 10^{-5}$~\cite{10.1093/ptep/ptaa104}), helium-4 ($^4\text{He}/{}^1\text{H} = (2.45 \pm 0.03)\times 10^{-1}$~\cite{10.1093/ptep/ptaa104}), and helium-3 ($^3\text{He}/\text{D} = (8.3 \pm 1.5)\times 10^{-1}$~\cite{GeissHe3}), the latter of which is only used to set an upper bound (c.f.~e.g.~\cite{Kawasaki:2004qu,GeissHe3,Ellis:2005ii}). Note that in this work we explicitly rederive these bounds in order to \textit{(i)} incorporate the most recent measurements of the light-element abundances, and \textit{(ii)} fully account for the mass dependence of the constraints (unlike for the CMB case where the very weak mass dependence allowed to rely on the previous literature).

Finally, let us note that the hadronic particles from the electroweak shower can also induce hadrodisintegration reactions, which can potentially lead to even stronger constraints (see e.g.~\cite{Kawasaki:2004qu,Jedamzik:2006xz,Jedamzik:2007qk}). However, we do not consider this mechanism here, meaning that our bounds are ultimately conservative, and could potentially be improved in future work.

\section{Results and discussion}\label{sec:dis}
Now that both the decay process and the cosmological probes used to constrain it have been outlined, we move on to the discussion of the resulting bounds. These are displayed in the $\tau_\phi-m_\phi$ plane in Fig.~\ref{fig:final} and in the $f_\phi-\tau_\phi$ plane in Fig.~\ref{fig:final_fphi}.

\subsection{Constraints from electromagnetic energy injection}

\begin{figure*}[t]
    \centering
    \includegraphics[width=0.48\columnwidth]{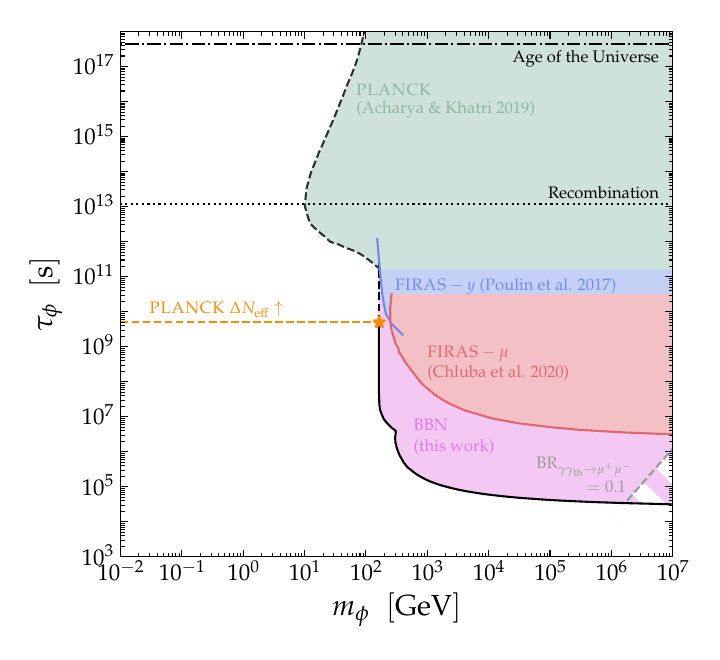}
    \includegraphics[width=0.48\columnwidth]{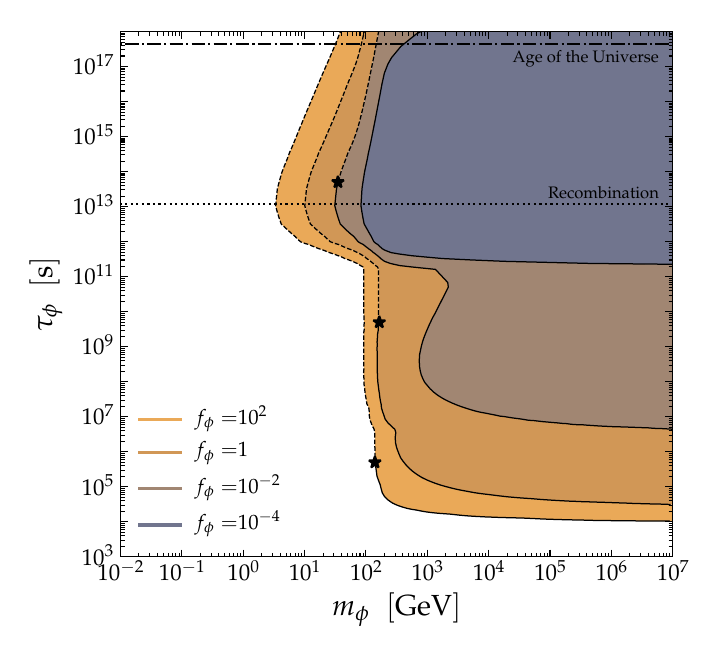}
    \caption{\textit{Left panel}: CMB and BBN constraints (see Sec.~\ref{sec:probes}) on the mass $m_\phi$ and lifetime $\tau_\phi$ of a massive particle decaying into electron neutrinos, $\phi\to\nu_e\bar{\nu}_e$ (see Sec.~\ref{sec:injection}) with a fractional abundance $f_\phi=\Omega_\phi/\Omega_{\rm DM}=1$. These constraints are derived solely from the injection of EM radiation following the decay, but we also show for reference the estimated region of parameter space excluded by $\Delta N_{\rm eff}$ constraints (above the orange dashed line/star, see Sec. \ref{sec: compl_const}). Overall, the solid/dashed black line indicates the cumulative constraint from the EM energy injection, which is dominant compared to the $\Delta N_\text{eff}$ constraint below (solid black) and subdominant above (dashed black) the star. \textit{Right panel}: Same as left panel but keeping only the best bound for $f_\phi=1$, showing in addition the corresponding best bounds for $f_\phi=10^{2,-2,-4}$. In all panels the time of recombination and the age of the Universe are shown for reference as black dotted and dashed-dotted horizontal lines, respectively.}
    \label{fig:final}
\end{figure*}

In the left panel of Fig.~\ref{fig:final}, we show the combination of the constraints resulting from CMB and BBN observations, obtained as discussed in Sec.~\ref{sec:probes} by using the relation\footnote{As a remark, if instead of using the correct red curve in Fig. \ref{fig:zeta} one was to use the PPPC prediction, the constraints displayed in Fig. \ref{fig:final} would be cut off at $m_\phi\simeq250$ GeV, which is where the PPPC predicts $\zeta_{\rm EM}$ to drop to zero.} presented in Fig.~\ref{fig:zeta}. Here, all bounds have been calculated for the arbitrary value $f_\phi=1$ (i.e.~for a $\phi$ abundance, which -- if it was not decaying -- would be equal today to the DM abundance). In this plot, the relatively small hatched area indicates the region where muon pairs can efficiently be produced and then partly decay back into neutrinos. As explained above, we neglect this effect and consequently the results are only approximate in this region, with the overall bound changing by at most 20\%.

Overall, we find that CMB anisotropy constraints (green) from Planck 2015 can probe mediator masses down to approximately 10$\,\mathrm{GeV}$ for lifetimes around the epoch of recombination (which we show as a dotted black line for reference). Note that these bounds are expected to improve~\cite{Lucca2019Synergy} with a Planck 2018 analysis and in the next decade with data from CMB-S4~\cite{Abazajian2016CMB, Abazajian2019CMB}, possibly reaching masses of the order of a fraction of a GeV. For lifetimes larger than roughly $10^{14}\,\mathrm{s}$, CMB anisotropy constraints on $m_\phi$ reduce linearly with $\tau_\phi$, and as soon as we consider lifetimes larger than the age of the Universe (indicated by the dashed-dotted black line) the cosmic-ray and high-energy neutrino experiments come into play. These latter bounds are summarised in e.g.~Fig.~2 of~\cite{Esmaili:2012us}.

\begin{figure*}[t]
    \centering
    \includegraphics[width=0.6\columnwidth]{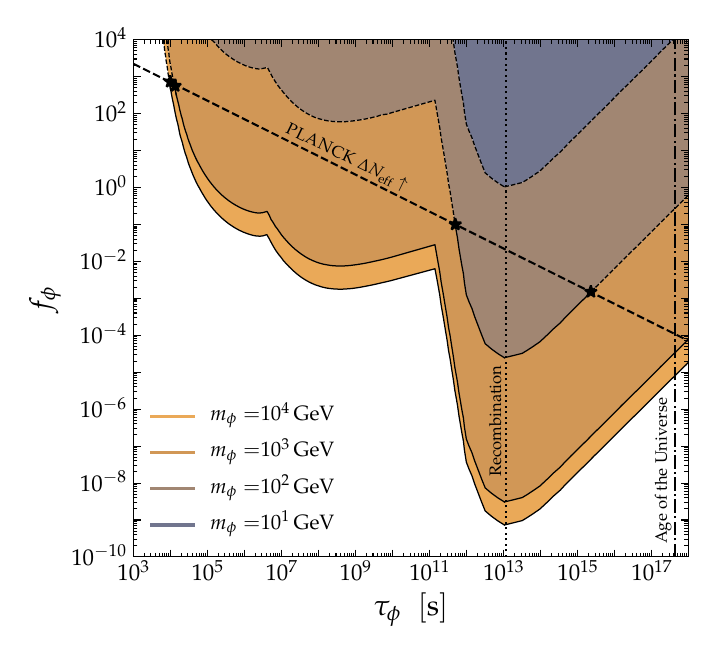}
    \caption{Same cumulative constraints as in the right panel of Fig.~\ref{fig:final}, but in the $f_\phi-\tau_\phi$ plane for different choices of the relic mass $m_\phi$. The diagonal dashed line indicates the estimated $\Delta N_{\rm eff}$ constraints discussed in Sec. \ref{sec: compl_const} (cf.~Eq.~\eqref{eq:Neff}).}
    \label{fig:final_fphi}
\end{figure*}

For lifetimes smaller than $10^{11}\,\mathrm{s}$ (or correspondingly redshifts of the order of $10^4$) these CMB limits drop very sharply and BBN constraints (purple) begin to dominate. In this case, masses above $\sim 100\,\mathrm{GeV}$ are excluded for lifetime larger than $\sim 10^5\,\mathrm{s}$. However, BBN photodisintegration does not constrain smaller lifetimes. Finally, although less constraining than BBN bounds, for the sake of generality we also include bounds from CMB SDs as (not) observed by FIRAS (blue and red). As already mentioned above, and as argued in a number of references (see e.g.~\cite{Lucca2019Synergy, Chluba:2020oip, Fu2020Unlocking}), these limits could be largely improved in the future. Already with PIXIE-like sensitivities~\cite{Kogut2011Primordial} future measurements of CMB SDs would improve on BBN bounds by orders of magnitude, potentially probing the $10-100\,\mathrm{GeV}$ range also for lifetimes smaller than recombination.

The right panel of Fig.~\ref{fig:final} shows how the bounds change when one varies the initial abundance of the decaying particle. For a value of $f_\phi$ equal to $10^2$, the bounds on $m_\phi$ improve by approximately half an order of magnitude with respect to the $f_\phi=1$ case of the left panel. This stems from the fact that, as explained in Sec.~\ref{sec:probes}, the bounds are on the product $f_\phi\zeta_{\rm EM}$. Thus, if one increases $f_\phi$, the amount of EM material produced, $\zeta_{\rm EM}$, has to decrease accordingly, and $\zeta_{\rm EM}$ scales as $m_\phi^{-4}$ for the masses of $\phi$ that apply here (that is to say below the $W/Z$ thresholds, see Fig.~\ref{fig:zeta}). Instead, as Fig.~\ref{fig:final} shows, for values of $f_\phi$ below unity the scaling is less trivial, as a result of the fact that in this case larger values of $\zeta_{\rm EM}$ are allowed, so that in Fig.~\ref{fig:zeta} masses of $\phi$ above the $W/Z$ threshold apply for the bounds (where the scaling of $\zeta_{\rm EM}$ in $m_\phi$ is different than below these thresholds). Note that as a result of the fact that above the thresholds the dependence of $\zeta_{\rm EM}$ on $m_\phi$ becomes gradually very weak, for a value of $f_\phi$ as small as $10^{-4}$, all masses below approximately 200$\,\mathrm{GeV}$ are allowed.

For the sake of completeness, in Fig.~\ref{fig:final_fphi} we also display the same cumulative bounds presented in the right panel of Fig.~\ref{fig:final}, but in the $f_\phi-\tau_\phi$ plane for different values of $m_\phi$. From the figure, it becomes evident that for masses above $\sim 1$ TeV the constraints on the fraction of decaying particles are only weakly sensitive to their mass. This is perfectly consistent with the flattening of the $\zeta_{\rm EM}$ scaling with respect to $m_\phi$ shown in Fig.~\ref{fig:zeta}. Furthermore, although already mentioned in the context of the left panel of Fig.~\ref{fig:final}, from Fig.~\ref{fig:final_fphi} one can also infer more clearly that for relic masses below 10 GeV every lifetime is allowed as long as the decaying particle constitutes only a sub-component of the DM content of the universe (i.e. if $f_\phi<1$).

We remark that for the derivation of these constraints we did not explicitly account for the impact of the decays on the expansion history of the universe, but focused solely on the bounds coming from the injection of EM energy. Nevertheless, one can estimate the region of parameter space where these corrections would in principle be relevant based on how the decays would affect the number of relativistic degrees of freedom $N_{\rm eff}$, as explained in the following section. We graphically show the resulting thresholds in Fig. \ref{fig:final} in form of stars for the different $f_\phi$ choices and highlight as dashed the sections of the bounds that might be affected by the aforementioned approximation. Similarly, in Fig. \ref{fig:final_fphi} the region of parameter space where the impact of the decays on $N_{\rm eff}$ becomes relevant lays above the diagonal dashed line (representing Eq.~\eqref{eq:Neff}). We remark, however, that the dashed regions would be anyway excluded by the $N_{\rm eff}$ constraints presented in the following section, and therefore our bounds are in any case conservative.

\subsection{The role of complementary constraints}\label{sec: compl_const}
Depending on the value assumed for $f_\phi$, other bounds that we do not consider here may apply. For instance, the larger the values of $f_\phi$ the more this particle, which becomes matter once non-relativistic, will dominate the energy density of the Universe until it decays. This can for instance modify the radiation/matter equality redshift $z_{\rm eq}$\,, which can be tightly constrained by CMB data \cite{Aghanim2018PlanckVI}. Furthermore, limits coming from the number of relativistic degrees of freedom that are allowed by BBN (already mentioned above) and CMB (from the production of additional neutrinos), see e.g.~\cite{Boyarsky:2021yoh}, could place significant bounds already for relatively small values of $f_\phi$. 

Focusing in particular on the latter effect, it is clear that a decay of the massive relic into neutrinos before recombination implies an extra amount of radiation at the recombination time and later (since the neutrinos do not thermalize with the photon thermal bath as long as the decay occurs after neutrino decoupling, i.e.~when $\tau_\phi \gtrsim 1$~s).  In fact, by solving the appropriate Boltzmann equations, we find that, for the extra particle injecting extra radiation when decaying into neutrinos, the current Planck measurement, $N_\text{eff} < 3.33$ at $95\%$ C.L.~\cite{Aghanim2018PlanckVI}, excludes all lifetimes above
\begin{equation}
    \tau_\phi f_\phi^2 \sim 5\times10^9\,\mathrm{s}\eqsp.
    \label{eq:Neff}
\end{equation}
We checked explicitly that very similar constraints can also be derived from Planck bounds on $z_{\rm eq}$. This explains why we do not consider values of the fractional abundance $f_\phi$ beyond $10^2$, as the relevant part of parameter space would be excluded anyway.

In Figs.~\ref{fig:final} and \ref{fig:final_fphi} we display these constraints in the form of stars, which indicate the value of $\tau_\phi$ above which $\Delta N_\text{eff}$ constraints become relevant. Also, in Fig.~\ref{fig:final_fphi} we additionally draw a dashed black line representing the relation from Eq.~\eqref{eq:Neff} (in the right panel of Fig.~\ref{fig:final}, the same equation would lead to one line for each value of $f_\phi$, which is why we omit them for the sake of graphical clearness). These thresholds serve both as an indicator for where the impact of the decay on the expansion history can be neglected, as mentioned above, and as an estimate of the region of parameter space excluded by $N_\text{eff}$ constraints. However, the relation presented in Eq.~\eqref{eq:Neff} is just an approximation and has to be taken as indicative for where the true constraints would lay. Indeed, the accurate determination of such bounds can be rather non-trivial (see e.g., Fig. 6 of \cite{Bringmann2018Converting} for a possible model-dependent representation of the maximal additional relativistic degrees of freedom that can be injected into the Universe as a function of time),  so that a more precise computation of these additional complementary constraints is left for future work.

As a final remark, the values of $f_\phi$ we consider are smaller than the ones one would obtain if, for instance, the decaying particle would thermalize with the SM thermal bath and decouple from it relativistically. For this extremely large abundance case we obtain limits on $m_\phi$ which are one order of magnitude more stringent than for the $f_\phi=10^2$ line in Fig.~\ref{fig:final}, but this possibility is anyway excluded by the $\Delta N_{\rm eff}$ constraints, unless the lifetime is quite short~\cite{Hufnagel:2018bjp,Depta:2020zbh,Hambye:2019tjt}. Values of $f_\phi$ such as those considered in Fig.~\ref{fig:final} can be obtained e.g.~for a non-relativistic decoupling or from a relativistic decoupling in a hidden sector thermal bath which would be colder than the SM thermal bath~\cite{Hambye2019Minimal, Hambye:2020lvy,Coy:2021ann}. The latter situation can apply to various self-interacting DM models with light mediators, see e.g.~\cite{Hambye2019Dark}, but for values of the light mediator mass below $\sim 100\,\mathrm{MeV}$ that are slightly smaller than the values of the masses probed in Fig.~\ref{fig:final}.

\section{Summary and conclusions}\label{sec: sum}

Many BSM setups involve new particles that could play a role in the early Universe. This includes DM scenarios involving dark sectors, where e.g.~the DM has its own new interactions mediated by new particles. The decay of such massive relics can significantly alter the thermal history of the Universe, especially if the lifetime of these particles is larger than the age of the Universe at the time of BBN, i.e.~$t\sim 1$~s. In particular, if a sizeable amount of the decay products ultimately end up in the form of EM radiation, these decays can have an effect on cosmological probes, such as on CMB anisotropies and SDs and on the BBN process.

Therefore, in order to avoid such stringent constraints, these unstable particles are often assumed to decay into neutrinos, as they do not interact as strongly with the thermal bath. However, even when this decay channel is assumed, i.e.~$\phi \rightarrow \nu_e \bar{\nu}_e$,  this process can have an effect on the aforementioned cosmological probes because they inevitably lead to the emission of on-shell or off-shell $W/Z$ gauge bosons, producing showers of electroweak and hadronic components, ultimately leading to an EM energy injection.

In this work, we determined the CMB and BBN constraints that this implies in terms of mass and lifetime of the decaying particle. We did that for lifetimes shorter than the age of the Universe and find new, stringent constraints that are presented in Fig.~\ref{fig:final}. They exclude any potential scenario where a decaying particle would have a lifetime larger than $\sim 10^4-10^7$~s and a mass above a few GeV to a few hundred GeV (depending on the lifetime and abundance of the decaying particle).

The fact that these constraints can go down to a few GeV masses stems from a proper calculation of the amount of the EM material that is produced just above the $W/Z$ threshold and below (from the emission of an off-shell gauge boson). We anticipate that these results for low masses (where \textsc{Pythia} gives vanishing results) could be relevant in other contexts too.

\section*{Acknowledgements}
We thank Marco Cirelli, Paul Frederik Depta, Nadege Iovine, and Filippo Sala for helpful discussions. MH and ML are supported by an F.R.S.-FNRS fellowship. This work is further supported by the \tquote{Probing dark matter with neutrinos} ULB-ARC convention, by the IISN convention 4.4503.15, and by the Excellence of Science (EoS) project No. 30820817 - be.h “The H boson gateway to physics beyond the Standard Model”. TH~further acknowledges  the support of the Institut Pascal at Université Paris-Saclay during the Paris-Saclay Astroparticle Symposium 2021, where part of this work has been done, and the related IN2P3 master project UCMN.


\bibliography{bibliography}

\providecommand{\href}[2]{#2}\begingroup\raggedright\begin{thebibliography}{10}

\bibitem{Feng:2009hw}
J.~L. Feng, M.~Kaplinghat and H.-B. Yu, \emph{{Halo Shape and Relic Density
  Exclusions of Sommerfeld-Enhanced Dark Matter Explanations of Cosmic Ray
  Excesses}},
  \href{http://dx.doi.org/10.1103/PhysRevLett.104.151301}{\emph{Phys. Rev.
  Lett.} {\bf 104} (2010) 151301}, [\href{https://arxiv.org/abs/0911.0422}{{\tt
  0911.0422}}].

\bibitem{Buckley:2009in}
M.~R. Buckley and P.~J. Fox, \emph{{Dark Matter Self-Interactions and Light
  Force Carriers}},
  \href{http://dx.doi.org/10.1103/PhysRevD.81.083522}{\emph{Phys. Rev. D} {\bf
  81} (2010) 083522}, [\href{https://arxiv.org/abs/0911.3898}{{\tt
  0911.3898}}].

\bibitem{Loeb:2010gj}
A.~Loeb and N.~Weiner, \emph{{Cores in Dwarf Galaxies from Dark Matter with a
  Yukawa Potential}},
  \href{http://dx.doi.org/10.1103/PhysRevLett.106.171302}{\emph{Phys. Rev.
  Lett.} {\bf 106} (2011) 171302}, [\href{https://arxiv.org/abs/1011.6374}{{\tt
  1011.6374}}].

\bibitem{Tulin:2013teo}
S.~Tulin, H.-B. Yu and K.~M. Zurek, \emph{{Beyond Collisionless Dark Matter:
  Particle Physics Dynamics for Dark Matter Halo Structure}},
  \href{http://dx.doi.org/10.1103/PhysRevD.87.115007}{\emph{Phys. Rev. D} {\bf
  87} (2013) 115007}, [\href{https://arxiv.org/abs/1302.3898}{{\tt
  1302.3898}}].

\bibitem{Bringmann:2016din}
T.~Bringmann, F.~Kahlhoefer, K.~Schmidt-Hoberg and P.~Walia, \emph{{Strong
  constraints on self-interacting dark matter with light mediators}},
  \href{http://dx.doi.org/10.1103/PhysRevLett.118.141802}{\emph{Phys. Rev.
  Lett.} {\bf 118} (2017) 141802},
  [\href{https://arxiv.org/abs/1612.00845}{{\tt 1612.00845}}].

\bibitem{Hambye:2019tjt}
T.~Hambye and L.~Vanderheyden, \emph{{Minimal self-interacting dark matter
  models with light mediator}},
  \href{http://dx.doi.org/10.1088/1475-7516/2020/05/001}{\emph{JCAP} {\bf 05}
  (2020) 001}, [\href{https://arxiv.org/abs/1912.11708}{{\tt 1912.11708}}].

\bibitem{Slatyer:2012yq}
T.~R. Slatyer, \emph{{Energy Injection And Absorption In The Cosmic Dark
  Ages}}, \href{http://dx.doi.org/10.1103/PhysRevD.87.123513}{\emph{Phys. Rev.
  D} {\bf 87} (2013) 123513}, [\href{https://arxiv.org/abs/1211.0283}{{\tt
  1211.0283}}].

\bibitem{Poulin2017Cosmological}
V.~Poulin, J.~Lesgourgues and P.~D. Serpico, \emph{{Cosmological constraints on
  exotic injection of electromagnetic energy}},
  \href{http://dx.doi.org/10.1088/1475-7516/2017/03/043}{\emph{JCAP} {\bf 1703}
  (2017) 043}, [\href{https://arxiv.org/abs/1610.10051}{{\tt 1610.10051}}].

\bibitem{Esmaili:2012us}
A.~Esmaili, A.~Ibarra and O.~L.~G. Peres, \emph{{Probing the stability of
  superheavy dark matter particles with high-energy neutrinos}},
  \href{http://dx.doi.org/10.1088/1475-7516/2012/11/034}{\emph{JCAP} {\bf 11}
  (2012) 034}, [\href{https://arxiv.org/abs/1205.5281}{{\tt 1205.5281}}].

\bibitem{ElAisati:2015ugc}
C.~El~Aisati, M.~Gustafsson and T.~Hambye, \emph{{New Search for Monochromatic
  Neutrinos from Dark Matter Decay}},
  \href{http://dx.doi.org/10.1103/PhysRevD.92.123515}{\emph{Phys. Rev. D} {\bf
  92} (2015) 123515}, [\href{https://arxiv.org/abs/1506.02657}{{\tt
  1506.02657}}].

\bibitem{Slatyer:2016qyl}
T.~R. Slatyer and C.-L. Wu, \emph{{General Constraints on Dark Matter Decay
  from the Cosmic Microwave Background}},
  \href{http://dx.doi.org/10.1103/PhysRevD.95.023010}{\emph{Phys. Rev. D} {\bf
  95} (2017) 023010}, [\href{https://arxiv.org/abs/1610.06933}{{\tt
  1610.06933}}].

\bibitem{Garcia-Cely:2017oco}
C.~Garcia-Cely and J.~Heeck, \emph{{Neutrino Lines from Majoron Dark Matter}},
  \href{http://dx.doi.org/10.1007/JHEP05(2017)102}{\emph{JHEP} {\bf 05} (2017)
  102}, [\href{https://arxiv.org/abs/1701.07209}{{\tt 1701.07209}}].

\bibitem{IceCube:2018tkk}
{\scshape IceCube} collaboration, M.~G. Aartsen et~al., \emph{{Search for
  neutrinos from decaying dark matter with IceCube}},
  \href{http://dx.doi.org/10.1140/epjc/s10052-018-6273-3}{\emph{Eur. Phys. J.
  C} {\bf 78} (2018) 831}, [\href{https://arxiv.org/abs/1804.03848}{{\tt
  1804.03848}}].

\bibitem{Bhattacharya:2019ucd}
A.~Bhattacharya, A.~Esmaili, S.~Palomares-Ruiz and I.~Sarcevic, \emph{{Update
  on decaying and annihilating heavy dark matter with the 6-year IceCube HESE
  data}}, \href{http://dx.doi.org/10.1088/1475-7516/2019/05/051}{\emph{JCAP}
  {\bf 05} (2019) 051}, [\href{https://arxiv.org/abs/1903.12623}{{\tt
  1903.12623}}].

\bibitem{Coy:2020wxp}
R.~Coy and T.~Hambye, \emph{{Neutrino lines from DM decay induced by high-scale
  seesaw interactions}},
  \href{http://dx.doi.org/10.1007/JHEP05(2021)101}{\emph{JHEP} {\bf 05} (2021)
  101}, [\href{https://arxiv.org/abs/2012.05276}{{\tt 2012.05276}}].

\bibitem{Coy:2021sse}
R.~Coy, A.~Gupta and T.~Hambye, \emph{{Seesaw neutrino determination of the
  dark matter relic density}},
  \href{http://dx.doi.org/10.1103/PhysRevD.104.083024}{\emph{Phys. Rev. D} {\bf
  104} (2021) 083024}, [\href{https://arxiv.org/abs/2104.00042}{{\tt
  2104.00042}}].

\bibitem{Cirelli2011PPPC}
M.~Cirelli et~al., \emph{{PPPC 4 DM ID: A Poor Particle Physicist Cookbook for
  Dark Matter Indirect Detection}},
  \href{http://dx.doi.org/10.1088/1475-7516/2012/10/E01,
  10.1088/1475-7516/2011/03/051}{\emph{JCAP} {\bf 1103} (2011) 051},
  [\href{https://arxiv.org/abs/1012.4515}{{\tt 1012.4515}}].

\bibitem{Queiroz:2016zwd}
F.~S. Queiroz, C.~E. Yaguna and C.~Weniger, \emph{{Gamma-ray Limits on Neutrino
  Lines}}, \href{http://dx.doi.org/10.1088/1475-7516/2016/05/050}{\emph{JCAP}
  {\bf 05} (2016) 050}, [\href{https://arxiv.org/abs/1602.05966}{{\tt
  1602.05966}}].

\bibitem{Stocker:2018avm}
P.~St\"ocker, M.~Kr\"amer, J.~Lesgourgues and V.~Poulin, \emph{{Exotic energy
  injection with ExoCLASS: Application to the Higgs portal model and
  evaporating black holes}},
  \href{http://dx.doi.org/10.1088/1475-7516/2018/03/018}{\emph{JCAP} {\bf 03}
  (2018) 018}, [\href{https://arxiv.org/abs/1801.01871}{{\tt 1801.01871}}].

\bibitem{Liu:2019bbm}
H.~Liu, G.~W. Ridgway and T.~R. Slatyer, \emph{{Code package for calculating
  modified cosmic ionization and thermal histories with dark matter and other
  exotic energy injections}},
  \href{http://dx.doi.org/10.1103/PhysRevD.101.023530}{\emph{Phys. Rev. D} {\bf
  101} (2020) 023530}, [\href{https://arxiv.org/abs/1904.09296}{{\tt
  1904.09296}}].

\bibitem{Sjostrand:2014zea}
T.~Sj\"ostrand, S.~Ask, J.~R. Christiansen, R.~Corke, N.~Desai, P.~Ilten
  et~al., \emph{{An introduction to PYTHIA 8.2}},
  \href{http://dx.doi.org/10.1016/j.cpc.2015.01.024}{\emph{Comput. Phys.
  Commun.} {\bf 191} (2015) 159--177},
  [\href{https://arxiv.org/abs/1410.3012}{{\tt 1410.3012}}].

\bibitem{Alloul:2013bka}
A.~Alloul, N.~D. Christensen, C.~Degrande, C.~Duhr and B.~Fuks,
  \emph{{FeynRules 2.0 - A complete toolbox for tree-level phenomenology}},
  \href{http://dx.doi.org/10.1016/j.cpc.2014.04.012}{\emph{Comput. Phys.
  Commun.} {\bf 185} (2014) 2250--2300},
  [\href{https://arxiv.org/abs/1310.1921}{{\tt 1310.1921}}].

\bibitem{Alwall:2011uj}
J.~Alwall, M.~Herquet, F.~Maltoni, O.~Mattelaer and T.~Stelzer, \emph{{MadGraph
  5 : Going Beyond}},
  \href{http://dx.doi.org/10.1007/JHEP06(2011)128}{\emph{JHEP} {\bf 06} (2011)
  128}, [\href{https://arxiv.org/abs/1106.0522}{{\tt 1106.0522}}].

\bibitem{Kawasaki:1994sc}
M.~Kawasaki and T.~Moroi, \emph{{Electromagnetic cascade in the early universe
  and its application to the big bang nucleosynthesis}},
  \href{http://dx.doi.org/10.1086/176324}{\emph{Astrophys. J.} {\bf 452} (1995)
  506}, [\href{https://arxiv.org/abs/astro-ph/9412055}{{\tt
  astro-ph/9412055}}].

\bibitem{Acharya:2020gfh}
S.~K. Acharya and R.~Khatri, \emph{{Constraints on $N_{\rm{eff}}$ of high
  energy non-thermal neutrino injections upto $z\sim 10^8$ from CMB spectral
  distortions and abundance of light elements}},
  \href{http://dx.doi.org/10.1088/1475-7516/2020/11/011}{\emph{JCAP} {\bf 11}
  (2020) 011}, [\href{https://arxiv.org/abs/2007.06596}{{\tt 2007.06596}}].

\bibitem{Kawasaki:2004qu}
M.~Kawasaki, K.~Kohri and T.~Moroi, \emph{{Big-Bang nucleosynthesis and
  hadronic decay of long-lived massive particles}},
  \href{http://dx.doi.org/10.1103/PhysRevD.71.083502}{\emph{Phys. Rev. D} {\bf
  71} (2005) 083502}, [\href{https://arxiv.org/abs/astro-ph/0408426}{{\tt
  astro-ph/0408426}}].

\bibitem{Jedamzik:2006xz}
K.~Jedamzik, \emph{{Big bang nucleosynthesis constraints on hadronically and
  electromagnetically decaying relic neutral particles}},
  \href{http://dx.doi.org/10.1103/PhysRevD.74.103509}{\emph{Phys. Rev. D} {\bf
  74} (2006) 103509}, [\href{https://arxiv.org/abs/hep-ph/0604251}{{\tt
  hep-ph/0604251}}].

\bibitem{Bauer:2020jay}
C.~W. Bauer, N.~L. Rodd and B.~R. Webber, \emph{{Dark matter spectra from the
  electroweak to the Planck scale}},
  \href{http://dx.doi.org/10.1007/JHEP06(2021)121}{\emph{JHEP} {\bf 06} (2021)
  121}, [\href{https://arxiv.org/abs/2007.15001}{{\tt 2007.15001}}].

\bibitem{Galli2013Systematic}
S.~Galli, T.~R. Slatyer, M.~Valdes and F.~Iocco, \emph{{Systematic
  Uncertainties In Constraining Dark Matter Annihilation From The Cosmic
  Microwave Background}},
  \href{http://dx.doi.org/10.1103/PhysRevD.88.063502}{\emph{Phys. Rev.} {\bf
  D88} (2013) 063502}, [\href{https://arxiv.org/abs/1306.0563}{{\tt
  1306.0563}}].

\bibitem{Slatyer2015Indirect}
T.~R. {Slatyer}, \emph{{Indirect dark matter signatures in the cosmic dark
  ages. II. Ionization, heating, and photon production from arbitrary energy
  injections}},
  \href{http://dx.doi.org/10.1103/PhysRevD.93.023521}{\emph{Phys.~Rev.} {\bf
  D93} (Jan., 2016) 023521}, [\href{https://arxiv.org/abs/1506.03812}{{\tt
  1506.03812}}].

\bibitem{Acharya2019CMB}
S.~K. Acharya and R.~Khatri, \emph{{CMB anisotropy and BBN constraints on
  pre-recombination decay of dark matter to visible particles}},
  \href{https://arxiv.org/abs/1910.06272}{{\tt 1910.06272}}.

\bibitem{Poulin2016Fresh}
V.~Poulin, P.~D. Serpico and J.~Lesgourgues, \emph{{A fresh look at linear
  cosmological constraints on a decaying dark matter component}},
  \href{https://arxiv.org/abs/1606.02073}{{\tt 1606.02073}}.

\bibitem{Ade2015PlanckXIII}
{\scshape Planck} collaboration, P.~A.~R. Ade et~al., \emph{{Planck 2015
  results. XIII. Cosmological parameters}},
  \href{http://dx.doi.org/10.1051/0004-6361/201525830}{\emph{Astron.
  Astrophys.} {\bf 594} (2016) A13},
  [\href{https://arxiv.org/abs/1502.01589}{{\tt 1502.01589}}].

\bibitem{Aghanim2018PlanckVI}
{\scshape Planck} collaboration, N.~Aghanim et~al., \emph{{Planck 2018 results.
  VI. Cosmological parameters}},
  \href{http://dx.doi.org/10.1051/0004-6361/201833910}{\emph{Astron.
  Astrophys.} {\bf 641} (2020) A6},
  [\href{https://arxiv.org/abs/1807.06209}{{\tt 1807.06209}}].

\bibitem{Lucca2019Synergy}
M.~Lucca, N.~Sch{\"o}neberg, D.~C. Hooper, J.~Lesgourgues and J.~Chluba,
  \emph{{The synergy between CMB spectral distortions and anisotropies}},
  \href{https://arxiv.org/abs/1910.04619}{{\tt 1910.04619}}.

\bibitem{Chluba2019Spectral}
J.~Chluba et~al., \emph{{Spectral Distortions of the CMB as a Probe of
  Inflation, Recombination, Structure Formation and Particle Physics}},
  {\emph{\baas} {\bf 51} (May, 2019) 184},
  [\href{https://arxiv.org/abs/1903.04218}{{\tt 1903.04218}}].

\bibitem{Chluba2019Voyage}
J.~Chluba et~al., \emph{{New Horizons in Cosmology with Spectral Distortions of
  the Cosmic Microwave Background}},
  \href{https://arxiv.org/abs/1909.01593}{{\tt 1909.01593}}.

\bibitem{Fu2020Unlocking}
H.~Fu, M.~Lucca, S.~Galli, E.~S. Battistelli, D.~C. Hooper, J.~Lesgourgues
  et~al., \emph{{Unlocking the synergy between CMB spectral distortions and
  anisotropies}},  \href{https://arxiv.org/abs/2006.12886}{{\tt 2006.12886}}.

\bibitem{Chluba2011Evolution}
J.~Chluba and R.~A. Sunyaev, \emph{{The evolution of CMB spectral distortions
  in the early Universe}},
  \href{http://dx.doi.org/10.1111/j.1365-2966.2011.19786.x}{\emph{Mon. Not.
  Roy. Astron. Soc.} {\bf 419} (2012) 1294--1314},
  [\href{https://arxiv.org/abs/1109.6552}{{\tt 1109.6552}}].

\bibitem{Chluba2016Which}
J.~Chluba, \emph{{Which spectral distortions does $\Lambda$CDM actually
  predict?}}, \href{http://dx.doi.org/10.1093/mnras/stw945}{\emph{Mon. Not.
  Roy. Astron. Soc.} {\bf 460} (2016) 227--239},
  [\href{https://arxiv.org/abs/1603.02496}{{\tt 1603.02496}}].

\bibitem{Hu1993thermalizationII}
W.~Hu and J.~Silk, \emph{Thermalization constraints and spectral distortions
  for massive unstable relic particles},
  \href{http://dx.doi.org/10.1103/PhysRevLett.70.2661}{\emph{Phys. Rev. Lett.}
  {\bf 70} (May, 1993) 2661--2664}.

\bibitem{Chluba2014Teasing}
J.~Chluba and D.~Jeong, \emph{{Teasing bits of information out of the CMB
  energy spectrum}}, \href{http://dx.doi.org/10.1093/mnras/stt2327}{\emph{Mon.
  Not. Roy. Astron. Soc.} {\bf 438} (2014) 2065--2082},
  [\href{https://arxiv.org/abs/1306.5751}{{\tt 1306.5751}}].

\bibitem{Acharya2019New}
S.~K. Acharya and R.~Khatri, \emph{{New CMB spectral distortion constraints on
  decaying dark matter with full evolution of electromagnetic cascades before
  recombination}},
  \href{http://dx.doi.org/10.1103/PhysRevD.99.123510}{\emph{Phys. Rev.} {\bf
  D99} (2019) 123510}, [\href{https://arxiv.org/abs/1903.04503}{{\tt
  1903.04503}}].

\bibitem{Chluba:2020oip}
J.~Chluba, A.~Ravenni and S.~K. Acharya, \emph{{Thermalization of large energy
  release in the early Universe}},
  \href{http://dx.doi.org/10.1093/mnras/staa2131}{\emph{Mon. Not. Roy. Astron.
  Soc.} {\bf 498} (2020) 959--980},
  [\href{https://arxiv.org/abs/2005.11325}{{\tt 2005.11325}}].

\bibitem{Acharya:2021zhq}
S.~K. Acharya and J.~Chluba, \emph{{CMB spectral distortions from continuous
  large energy release}},  \href{https://arxiv.org/abs/2112.06699}{{\tt
  2112.06699}}.

\bibitem{Kogut2011Primordial}
A.~Kogut et~al., \emph{{The Primordial Inflation Explorer (PIXIE): a nulling
  polarimeter for cosmic microwave background observations}},
  \href{http://dx.doi.org/10.1088/1475-7516/2011/07/025}{\emph{Journal of
  Cosmology and Astro-Particle Physics} {\bf 2011} (Jul, 2011) 025},
  [\href{https://arxiv.org/abs/1105.2044}{{\tt 1105.2044}}].

\bibitem{Andre2014Prism}
{\scshape PRISM} collaboration, P.~André et~al., \emph{{PRISM (Polarized
  Radiation Imaging and Spectroscopy Mission): An Extended White Paper}},
  \href{http://dx.doi.org/10.1088/1475-7516/2014/02/006}{\emph{JCAP} {\bf 1402}
  (2014) 006}, [\href{https://arxiv.org/abs/1310.1554}{{\tt 1310.1554}}].

\bibitem{Kogut2019CMB}
A.~Kogut, M.~Abitbol, J.~Chluba, J.~Delabrouille, D.~Fixsen, J.~Hill et~al.,
  \emph{{CMB Spectral Distortions: Status and Prospects}},
  \href{https://arxiv.org/abs/1907.13195}{{\tt 1907.13195}}.

\bibitem{Mather1994Measurement}
J.~C. Mather et~al., \emph{{Measurement of the cosmic microwave background
  spectrum by the COBE FIRAS instrument}},
  \href{http://dx.doi.org/10.1086/173574}{\emph{\apj} {\bf 420} (Jan., 1994)
  439--444}.

\bibitem{Fixsen1996Cosmic}
D.~J. Fixsen, E.~S. Cheng, J.~M. Gales, J.~C. Mather, R.~A. Shafer and E.~L.
  Wright, \emph{{The Cosmic Microwave Background spectrum from the full COBE
  FIRAS data set}}, \href{http://dx.doi.org/10.1086/178173}{\emph{Astrophys.
  J.} {\bf 473} (1996) 576},
  [\href{https://arxiv.org/abs/astro-ph/9605054}{{\tt astro-ph/9605054}}].

\bibitem{Cyburt:2015mya}
R.~H. Cyburt, B.~D. Fields, K.~A. Olive and T.-H. Yeh, \emph{{Big Bang
  Nucleosynthesis: 2015}},
  \href{http://dx.doi.org/10.1103/RevModPhys.88.015004}{\emph{Rev. Mod. Phys.}
  {\bf 88} (2016) 015004}, [\href{https://arxiv.org/abs/1505.01076}{{\tt
  1505.01076}}].

\bibitem{10.1093/ptep/ptaa104}
PDG, \emph{{Review of Particle Physics}},
  \href{http://dx.doi.org/10.1093/ptep/ptaa104}{\emph{Progress of Theoretical
  and Experimental Physics} {\bf 2020} (08, 2020) },
  [\href{https://arxiv.org/abs/https://academic.oup.com/ptep/article-pdf/2020/8/083C01/34673722/ptaa104.pdf}{{\tt
  https://academic.oup.com/ptep/article-pdf/2020/8/083C01/34673722/ptaa104.pdf}}].

\bibitem{Depta:2019lbe}
P.~F. Depta, M.~Hufnagel, K.~Schmidt-Hoberg and S.~Wild, \emph{{BBN constraints
  on the annihilation of MeV-scale dark matter}},
  \href{http://dx.doi.org/10.1088/1475-7516/2019/04/029}{\emph{JCAP} {\bf 04}
  (2019) 029}, [\href{https://arxiv.org/abs/1901.06944}{{\tt 1901.06944}}].

\bibitem{Depta:2020zbh}
P.~F. Depta, M.~Hufnagel and K.~Schmidt-Hoberg, \emph{{Updated BBN constraints
  on electromagnetic decays of MeV-scale particles}},
  \href{http://dx.doi.org/10.1088/1475-7516/2021/04/011}{\emph{JCAP} {\bf 04}
  (2021) 011}, [\href{https://arxiv.org/abs/2011.06519}{{\tt 2011.06519}}].

\bibitem{Hufnagel:2017dgo}
M.~Hufnagel, K.~Schmidt-Hoberg and S.~Wild, \emph{{BBN constraints on MeV-scale
  dark sectors. Part I. Sterile decays}},
  \href{http://dx.doi.org/10.1088/1475-7516/2018/02/044}{\emph{JCAP} {\bf 02}
  (2018) 044}, [\href{https://arxiv.org/abs/1712.03972}{{\tt 1712.03972}}].

\bibitem{Hufnagel:2018bjp}
M.~Hufnagel, K.~Schmidt-Hoberg and S.~Wild, \emph{{BBN constraints on MeV-scale
  dark sectors. Part II. Electromagnetic decays}},
  \href{http://dx.doi.org/10.1088/1475-7516/2018/11/032}{\emph{JCAP} {\bf 11}
  (2018) 032}, [\href{https://arxiv.org/abs/1808.09324}{{\tt 1808.09324}}].

\bibitem{Depta:2020wmr}
P.~F. Depta, M.~Hufnagel and K.~Schmidt-Hoberg, \emph{{Robust cosmological
  constraints on axion-like particles}},
  \href{http://dx.doi.org/10.1088/1475-7516/2020/05/009}{\emph{JCAP} {\bf 05}
  (2020) 009}, [\href{https://arxiv.org/abs/2002.08370}{{\tt 2002.08370}}].

\bibitem{Kawasaki:2020qxm}
M.~Kawasaki, K.~Kohri, T.~Moroi, K.~Murai and H.~Murayama, \emph{{Big-bang
  nucleosynthesis with sub-GeV massive decaying particles}},
  \href{http://dx.doi.org/10.1088/1475-7516/2020/12/048}{\emph{JCAP} {\bf 12}
  (2020) 048}, [\href{https://arxiv.org/abs/2006.14803}{{\tt 2006.14803}}].

\bibitem{Poulin:2015opa}
V.~Poulin and P.~D. Serpico, \emph{{Nonuniversal BBN bounds on
  electromagnetically decaying particles}},
  \href{http://dx.doi.org/10.1103/PhysRevD.91.103007}{\emph{Phys. Rev. D} {\bf
  91} (2015) 103007}, [\href{https://arxiv.org/abs/1503.04852}{{\tt
  1503.04852}}].

\bibitem{Poulin:2015woa}
V.~Poulin and P.~D. Serpico, \emph{{Loophole to the Universal Photon Spectrum
  in Electromagnetic Cascades and Application to the Cosmological Lithium
  Problem}},
  \href{http://dx.doi.org/10.1103/PhysRevLett.114.091101}{\emph{Phys. Rev.
  Lett.} {\bf 114} (2015) 091101},
  [\href{https://arxiv.org/abs/1502.01250}{{\tt 1502.01250}}].

\bibitem{Forestell:2018txr}
L.~Forestell, D.~E. Morrissey and G.~White, \emph{{Limits from BBN on Light
  Electromagnetic Decays}},
  \href{http://dx.doi.org/10.1007/JHEP01(2019)074}{\emph{JHEP} {\bf 01} (2019)
  074}, [\href{https://arxiv.org/abs/1809.01179}{{\tt 1809.01179}}].

\bibitem{Jedamzik:2009uy}
K.~Jedamzik and M.~Pospelov, \emph{{Big Bang Nucleosynthesis and Particle Dark
  Matter}}, \href{http://dx.doi.org/10.1088/1367-2630/11/10/105028}{\emph{New
  J. Phys.} {\bf 11} (2009) 105028},
  [\href{https://arxiv.org/abs/0906.2087}{{\tt 0906.2087}}].

\bibitem{Boyarsky:2021yoh}
A.~Boyarsky, M.~Ovchynnikov, N.~Sabti and V.~Syvolap, \emph{{When feebly
  interacting massive particles decay into neutrinos: The Neff story}},
  \href{http://dx.doi.org/10.1103/PhysRevD.104.035006}{\emph{Phys. Rev. D} {\bf
  104} (2021) 035006}, [\href{https://arxiv.org/abs/2103.09831}{{\tt
  2103.09831}}].

\bibitem{Depta:2020mhj}
P.~F. Depta, M.~Hufnagel and K.~Schmidt-Hoberg, \emph{{ACROPOLIS: A generiC
  fRamework fOr Photodisintegration Of LIght elementS}},
  \href{http://dx.doi.org/10.1088/1475-7516/2021/03/061}{\emph{JCAP} {\bf 03}
  (2021) 061}, [\href{https://arxiv.org/abs/2011.06518}{{\tt 2011.06518}}].

\bibitem{Gandhi:1998ri}
R.~Gandhi, C.~Quigg, M.~H. Reno and I.~Sarcevic, \emph{{Neutrino interactions
  at ultrahigh-energies}},
  \href{http://dx.doi.org/10.1103/PhysRevD.58.093009}{\emph{Phys. Rev. D} {\bf
  58} (1998) 093009}, [\href{https://arxiv.org/abs/hep-ph/9807264}{{\tt
  hep-ph/9807264}}].

\bibitem{GeissHe3}
J.~Geiss and G.~Gloeckler, \emph{Isotopic composition of h, he and ne in the
  protosolar cloud},
  \href{http://dx.doi.org/10.1023/A:1024651232758}{\emph{Space Science Reviews}
  {\bf 106} (04, 2003) }.

\bibitem{Ellis:2005ii}
J.~R. Ellis, K.~A. Olive and E.~Vangioni, \emph{{Effects of unstable particles
  on light-element abundances: Lithium versus deuterium and He-3}},
  \href{http://dx.doi.org/10.1016/j.physletb.2005.05.066}{\emph{Phys. Lett. B}
  {\bf 619} (2005) 30--42}, [\href{https://arxiv.org/abs/astro-ph/0503023}{{\tt
  astro-ph/0503023}}].

\bibitem{Jedamzik:2007qk}
K.~Jedamzik, \emph{{Bounds on long-lived charged massive particles from Big
  Bang nucleosynthesis}},
  \href{http://dx.doi.org/10.1088/1475-7516/2008/03/008}{\emph{JCAP} {\bf 03}
  (2008) 008}, [\href{https://arxiv.org/abs/0710.5153}{{\tt 0710.5153}}].

\bibitem{Abazajian2016CMB}
{\scshape CMB-S4} collaboration, K.~N. Abazajian et~al., \emph{{CMB-S4 Science
  Book, First Edition}},  \href{https://arxiv.org/abs/1610.02743}{{\tt
  1610.02743}}.

\bibitem{Abazajian2019CMB}
K.~Abazajian et~al., \emph{{CMB-S4 Science Case, Reference Design, and Project
  Plan}},  \href{https://arxiv.org/abs/1907.04473}{{\tt 1907.04473}}.

\bibitem{Bringmann2018Converting}
T.~Bringmann, F.~Kahlhoefer, K.~Schmidt-Hoberg and P.~Walia, \emph{{Converting
  nonrelativistic dark matter to radiation}},
  \href{https://arxiv.org/abs/1803.03644}{{\tt 1803.03644}}.

\bibitem{Hambye2019Minimal}
T.~Hambye and L.~Vanderheyden, \emph{{Minimal self-interacting dark matter
  models with light mediator}},  \href{https://arxiv.org/abs/1912.11708}{{\tt
  1912.11708}}.

\bibitem{Hambye:2020lvy}
T.~Hambye, M.~Lucca and L.~Vanderheyden, \emph{{Dark matter as a heavy thermal
  hot relic}},
  \href{http://dx.doi.org/10.1016/j.physletb.2020.135553}{\emph{Phys. Lett. B}
  {\bf 807} (2020) 135553}, [\href{https://arxiv.org/abs/2003.04936}{{\tt
  2003.04936}}].

\bibitem{Coy:2021ann}
R.~Coy, T.~Hambye, M.~H.~G. Tytgat and L.~Vanderheyden, \emph{{Domain of
  thermal dark matter candidates}},
  \href{http://dx.doi.org/10.1103/PhysRevD.104.055021}{\emph{Phys. Rev. D} {\bf
  104} (2021) 055021}, [\href{https://arxiv.org/abs/2105.01263}{{\tt
  2105.01263}}].

\bibitem{Hambye2019Dark}
T.~Hambye, M.~H.~G. Tytgat, J.~Vandecasteele and L.~Vanderheyden, \emph{{Dark
  matter from dark photons: a taxonomy of dark matter production}},
  \href{http://dx.doi.org/10.1103/PhysRevD.100.095018}{\emph{Phys. Rev.} {\bf
  D100} (2019) 095018}, [\href{https://arxiv.org/abs/1908.09864}{{\tt
  1908.09864}}].

\end{thebibliography}\endgroup
\bibliographystyle{JHEP}

\end{document}